\documentclass[aps,pre,twocolumn,showpacs,preprintnumbers,superscriptaddress]{revtex4}

\usepackage{graphicx}
\usepackage{epsf} 
\usepackage{dcolumn}
\usepackage{bm}
\usepackage{amssymb}
\usepackage{amsmath}

\begin{document}

\title{Scaling Properties of Growing Noninfinitesimal
Perturbations in Space-Time Chaos}

\author{Juan M. L{\'o}pez} \email{lopez@ifca.unican.es}

\affiliation{Instituto de F\'{\i}sica de Cantabria (CSIC--UC),
E-39005 Santander, Spain}

\author{Cristina Primo}

\affiliation{Instituto de F\'{\i}sica de Cantabria (CSIC--UC),
E-39005 Santander, Spain}

\affiliation{Instituto Nacional de Meteorolog{\'\i}a CMT/CAS,
Santander, Spain}

\affiliation{Departamento de Matem\'atica Aplicada, Universidad de
Cantabria, Avda. Los Castros, E-39005 Santander, Spain}

\author{Miguel A. Rodr{\'\i}guez} 

\affiliation{Instituto de F\'{\i}sica de Cantabria (CSIC--UC),
E-39005 Santander, Spain}

\author{Ivan G.\ Szendro} 

\affiliation{Instituto de F\'{\i}sica de Cantabria (CSIC--UC),
E-39005 Santander, Spain}

\affiliation{Departamento de F{\'\i}sica Moderna, Universidad de
Cantabria, Avda. Los Castros, E-39005 Santander, Spain}
\date{\today}

\begin{abstract}
We study the spatiotemporal dynamics of random spatially
distributed noninfinitesimal perturbations in one-dimensional
chaotic extended systems. We find that an initial perturbation of
finite size $\epsilon_0$ grows in time obeying the tangent space
dynamic equations (Lyapunov vectors) up to a characteristic time
$t_{\times}(\epsilon_0) \sim b - (1/\lambda_{max}) \ln
(\epsilon_0)$, where $\lambda_{max}$ is the largest Lyapunov
exponent and $b$ is a constant. For times $t < t_{\times}$
perturbations exhibit spatial correlations up to a typical
distance $\xi \sim t^z$. For times larger than $t_{\times}$ finite
perturbations are no longer described by tangent space equations,
memory of spatial correlations is progressively destroyed and
perturbations become spatiotemporal white noise. We are able to
explain these results by mapping the problem to the
Kardar-Parisi-Zhang universality class of surface growth.
\end{abstract}


\maketitle

\section{introduction}
A standard tool for studying chaotic behavior in dynamical systems
is the computation of the characteristic Lyapunov exponents, which
measure the typical exponential growth rate of an infinitesimal
disturbance \cite{schuster,ott}. The characteristic Lyapunov
exponents in extended systems are defined in a similar way as
their low-dimensional counterpart and can be calculated from the
linearization of the equations of motion \cite{bohr,boffetta}. The
main point is that the growth of an infinitesimal perturbation is
described by the linear equations for the tangent space, the
so-called Lyapunov vectors (see below). However, for many
practical purposes the dynamics of infinitesimal perturbations may
be irrelevant as indicators of, for instance, the predictability
time. Indeed, in many realistic situations the error in the
initial conditions is finite. The important fact is that the
evolution of finite errors is not confined to the tangent space,
as defined by the growth of linearized perturbations, but is
controlled by the complete non-linear dynamics. A good example
with important practical application occurs in weather
forecasting, where one deals with the whole Earth's atmosphere--
an extremely high dimensional system in which initial conditions
can be determined only with limited accuracy. The effects of
finite perturbations have recently been studied in the context of
fully developed turbulence \cite{aurell}. In order to deal with
realistic perturbations, the concept of finite size Lyapunov
exponents has been found to be useful to analyze predictability in
high-dimensional chaotic systems \cite{boffetta,aurell}.

In this paper we study the dynamics of random spatially
distributed finite-size errors in chaotic extended systems and
focus on their propagation dynamics. We argue that, after a
suitable transformation of variables, the dynamics of finite
perturbations can be interpreted as a kinetic roughening process
in the Kardar-Parisi-Zhang (KPZ) universality class \cite{kpz}. We
find that, due to the finiteness of the initial error, there is a
characteristic time scale $t_{\times}(\epsilon_0) \sim b -
(1/\lambda_{max}) \ln (\epsilon_0)$, where $\lambda_{max}$ is the
largest Lyapunov exponent, $\epsilon_0$ is a measure of the
initial size of the perturbation, and $b$ is a constant. For times
$t < t_{\times}$ the dynamic evolution of a finite perturbation is
governed by the Lyapunov vector. In this regime, finite
perturbations become spatially correlated up to a typical length
scale $\xi \sim t_{\times}^z$, where $z$ is the dynamic exponent
of the KPZ problem ($z=3/2$ for one dimensional systems). However,
for times $t > t_{\times}$ any finite perturbation leaves the
tangent space and is no longer described by the Lyapunov vectors.
In this late regime, memory of spatial correlations is
progressively destroyed and perturbations actually become white
noise in space and time. Our approach provides new tools for
studying chaotic extended systems by allowing to fully describe
the spread of correlations, to estimate the spatial extend of
correlations of the chaotic field and to measure the effective
number of degrees of freedom.

We exemplify our results by means of numerical simulations of
coupled map lattices in one dimension, which are simple model
systems exhibiting space-time chaos and convenient as far as the
computing time is concerned. We consider a coupled map array
consisting of $L$ chaotic oscillators given by
\begin{eqnarray}
u(x,t+1)= \nu \,f(u(x+1,t))\,+
\nonumber\\
\nu\, f(u(x-1,t))\,+\,(1-2\,\nu)\,f(u(x,t)) \label{cml}
\end{eqnarray}
where $x = 1, 2, \ldots, L$, $f(u)$ is a chaotic map, $\nu$ is the
coupling constant, and periodic boundary conditions are imposed.
We have fixed the coupling to $\nu = 1/3$ in all the simulations
presented in this paper. We have carried out simulations for two
different choices of the map, the chaotic logistic map $f(u) =
4u(1-u)$, $0 \le u \le 1$ and the tent map $f(u) = 1 - 2\vert
u-1/2\vert$, $0 \le u \le 1$. For the sake of brevity, all the
results we present below correspond to coupled logistic maps, but
similar results were obtained for the tent map.

The dynamics of {\em infinitesimal} perturbations of a turbulent
state $u^0(x,t)$ can be studied by linearizing the evolution
equation Eq.\ (\ref{cml}). This leads to the tangent space
equations or Lyapunov vector $\delta u(x,t)$ evolution equation
\begin{eqnarray}
\delta u(x,t+1)= \nu \,f'[u(x+1,t)]\,\delta u(x+1,t)\, +
\nonumber\\
\nu\, f'[u(x-1,t)]\,\delta u(x-1,t)\, +
\nonumber\\
(1-2\,\nu)\,f'[u(x,t)]\,\delta u(x,t)\, + O[(\delta u)^2]
\label{lyap-cml}
\end{eqnarray}
where $f'[u(x,t)] = df(y)/dy \vert_{u(x,t)}$. This implies to
solve simultaneously the field $u(x,t)$ evolution equation
(\ref{cml}). The Lyapunov vector is defined in the linear
approximation, when higher order corrections $O[(\delta u)^2]$ are
neglected.

The analysis of infinitesimal perturbations allows to compute
several indicators characterizing the chaotic system, including
the whole spectrum of Lyapunov exponents and investigate how this
depends on system size \cite{boffetta,bohr,livi}. However, as
explained above, there are indeed many situations where the
Lyapunov analysis has no relevance due, for instance, to the
finite nature of the initial errors.

\section{Scaling of finite-size perturbations}
Let us now consider the evolution of random finite-size perturbed
trajectories in our model system (\ref{cml}). Given an initial
condition $u^0(x,0)$, the solution $u^0(x,t)$ is univocally
determined by computing Eq.\ (\ref{cml}) for a number $t$ of time
steps. This is our reference trajectory and we are interested in
the evolution of finite perturbations around that reference
solution. For reasons that will become clear below, we find
convenient to introduce now what we call the {\em amplitude
factor} $\epsilon(t)$ as the spatial geometrical mean value of a
perturbation:
\begin{equation}
\epsilon(t) \equiv \prod_{x=1}^L \vert \delta u(x,t)\vert^{1/L}.
\label{amplitude}
\end{equation}
As we shall see below, the amplitude factor turns out to be a very
important quantity which contains the information about the
dominant exponential growth rate.

Since we are interested here in the propagation of real
(non-infinitesimal) errors, we should avoid linearization of Eq.\
(\ref{cml}). Instead, we compute the trajectories generated by
iterating (\ref{cml}) for an ensemble of initial conditions
$u(x,0) = u^0(x,0)+\delta u(x,0)$, where $\delta u(x,0)$ is a
random finite perturbation of initial amplitude $\epsilon(t=0) =
\epsilon_0$. For each iteration of the lattice (\ref{cml}) the
difference $\delta u(x,t)=u(x,t)-u^0(x,t )$ between the reference
trajectory and every one of the perturbed solutions is calculated.
Although the disturbances are initially random and uncorrelated in
space, as time goes by, they grow and get spatially correlated.
The statistical fluctuations of errors can be characterized by
studying the ensemble of finite perturbations $\{\delta
u_{n}(x,t)\}_{n=1}^{N}$, which correspond to $N$ independent
realizations of the initial perturbation.

\begin{figure}
\centerline{\epsfxsize=7.5cm \epsfbox{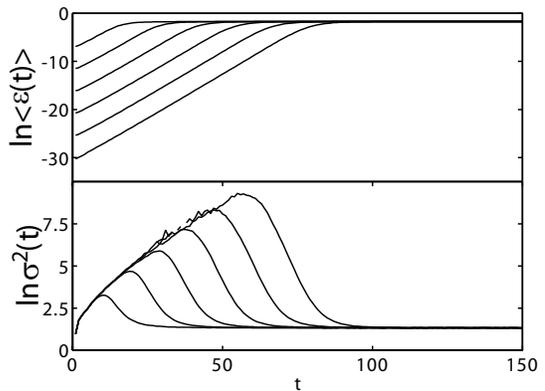}}
\caption{Numerical results for the propagation of finite-size
errors in coupled logistic maps. Upper panel shows the ensemble
averaged amplitude factor $\langle \epsilon(t) \rangle$ {\it vs.}
time for perturbations starting with initial amplitudes of
$\epsilon_0 = 10^{-3}, 10^{-5}, 10^{-7}, 10^{-9}, 10^{-11}$ and
$10^{-13}$ (from top to bottom) in 1D lattices of $L=1024$ sites.
Results were averaged over 600 different initial conditions. Lower
panel shows the variance $\sigma^2(t)$ for the same initial
perturbations as before and $\epsilon_0$ decreasing from $10^{-3}$
(leftmost curve) to $10^{-13}$ (rightmost curve).}
\end{figure}

In Fig.\ 1 (upper panel) we plot $\ln \langle \epsilon(t) \rangle$
{\it vs.} time for different values of the initial perturbation
amplitude $\epsilon_0$, where $\langle \cdots \rangle$ stands for
average over realizations of the initial perturbation. One can
immediately see that there exists a characteristic time scale
$t_{\times}(\epsilon_0)$ such that for times $t <
t_{\times}(\epsilon_0)$ the average amplitude factor grows
exponentially in time $\epsilon (t) \approx \epsilon_{0}
\exp(\lambda t)$, where $\lambda = 0.343 \pm 0.005$. We
demonstrate below that $\lambda$ indeed corresponds to the maximal
Lyapunov exponent. For longer times, $t
> t_{\times}(\epsilon_0)$ the amplitude factor saturates to a
constant value. Both, the saturation constant and the maximal
Lyapunov exponent, are independent of the initial perturbation
size $\epsilon_0$. However, the saturation times
$t_{\times}(\epsilon_0)$ increase as the size of the initial
perturbation $\epsilon_0$ becomes smaller. The characteristic time
scale $t_{\times}$ corresponds to the crossover time at which the
dynamics of a finite size perturbation depart from the linear
evolution ({\it i.e.} the Lyapunov vectors) given by Eq.\
(\ref{lyap-cml}). This crossover occurs because Lyapunov vectors
describe only the behavior of strictly infinitesimal
perturbations. One can estimate $t_{\times}$ roughly as the time
at which $\epsilon(t)$ reaches some $\delta$, so that the higher
order terms $O[(\delta u)^2]$ cannot be neglected in the evolution
equation (\ref{lyap-cml}). This crossover takes place at a typical
time $t_{\times}(\epsilon_0) \sim (1/\lambda) \ln(\delta) -
(1/\lambda) \ln(\epsilon_0)$. Therefore, for times $t
> t_{\times}(\epsilon_0)$ nonlinear corrections, due to finiteness
of the initial perturbation, come into play and drive errors out
of the tangent space. From then on, the linear approximation
cannot describe the evolution of errors. One then expects that
$t_{\times}(\epsilon_0) \to \infty$ as $\epsilon_0 \to 0$.

Besides exponential growth, spatial correlations are dynamically
generated during the evolution of the perturbation. Correlations
contain information about the sub-leading Lyapunov exponents and
thus also contribute to the perturbation size growth. The
important role of correlations can be better realized after
subtraction of the dominant exponential growth component given by
$\epsilon(t)$. We find that a very useful indicator is given by
the {\em reduced} perturbations $\delta r(x,t)$ that we define as
\begin{equation}
\delta r(x,t)=\frac{\delta u(x,t)}{\epsilon(t)}, \label{reduced}
\end{equation}
where the dominant exponential growth is globally removed and one
is left with the effect of correlations. For random initial
errors, {\it i.e.} random spatially distributed initial
conditions, the average $\langle \overline{\delta r(x,t) \rangle}$
vanishes, where $\langle \cdots \rangle$ stands for average over
realizations of the initial perturbation and the over bar is a
spatial average. Thus, statistical fluctuations of the reduced
perturbations are measured by the variance $\sigma^2(t) = \langle
\overline{\delta r(x,t)^2}\rangle$. As we shall see below,
$\sigma(t)$ gives information about the growth of perturbations
due solely to correlations.

In Fig.\ 1 (lower panel) we show our numerical results for the
variance of the reduced perturbations, Eq.(\ref{reduced}). The
variance $\sigma^2(t)$ grows as $\sigma^2(t)\sim \exp(t^{2\beta})$
for times $t < t_{\times}$, {\it i.e.} before saturation occurs
due to finiteness of the initial disturbance. This scaling
behavior is better seen in Fig.\ 2 where we plot in log-log scale
$\ln(\sigma^2)$ {\it vs.} time and the exponent $\beta = 0.30 \pm
0.05$ (see Fig.\ 2). As before, this rapid growth occurs for times
$t < t_{\times}$ when the dynamics of disturbances are well
described by the Lyapunov vectors. One can determine the crossover
time $t_{\times}$ from Fig.\ 1, either by measuring the saturation
times of $\epsilon(t)$ (upper panel) or by measuring the time
shift of the maxima of $\sigma^2(t)$ (lower panel). Results are
shown in the inset of Fig.\ 2, where a straight line fit of the
data leads to $t_{\times} (\epsilon_0) = b -(1/\lambda)
\ln(\epsilon_0)$ with $b$ a constant. The slope of the fit
$1/\lambda = 2.86 \pm 0.05$ is in excellent agreement with our
previous determination of $\lambda \approx 0.34$ from the
exponential growth of the amplitude factor $\epsilon(t) \sim
\exp(\lambda t)$.

\begin{figure}
\centerline{\epsfxsize=7.5cm \epsfbox{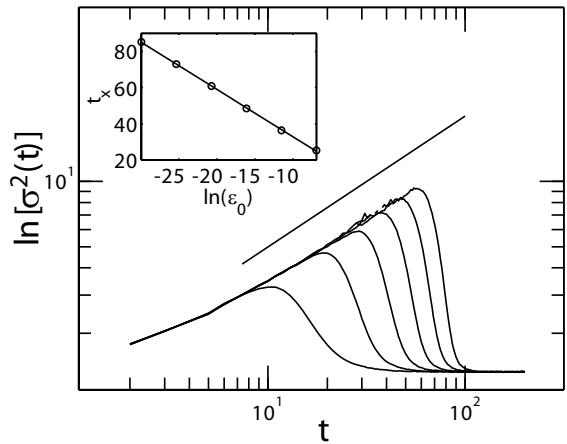}}
\caption{The same numerical results for the variance $\sigma^2(t)$
as those in Fig.\ 1 are plotted in the main panel in log-log scale
to show the power-law behavior $\ln \sigma^2(t) \sim t^{2\beta}$.
The straight line has a slope $0.60$ and is plotted to guide the
eye. The inset shows the mean-square fit of the characteristic
time scale $t_{\times}(\epsilon_0)$ obtained from the saturation
times of the amplitude factor $\epsilon(t)$. The slope of the
straight line is $1/\lambda = 2.86(5)$.}
\end{figure}

The magnitude and extent of spatial correlations can be measured
by using the site--site correlation function $G(l,t) = \langle
\overline{\delta r(x_0,t) \delta r(x_0+l,t)}\rangle/\langle
\overline{\delta r(x_0,t)^2}\rangle$. The magnitude of spatial
correlations increases in time $G(l,t_1) < G(l,t_2)$ if $t_1 <
t_2$ for times $t_1, t_2 < t_{\times}$. However, as shown in Fig.\
3, correlations become progressively smaller for times $t >
t_{\times}$. This again demonstrates that the effect of having
initially finite perturbations is to introduce a characteristic
time $t_{\times}(\epsilon_0)$ marking the typical time it takes
for the system to depart from tangent space (with the building-up
of correlations) to non-linear evolution (uncorrelated errors).
For times larger than $t_{\times}$ the perturbation spatial
correlation are progressively destroyed, as can be seen from the
decay of the correlations $G(l,t)$ in Fig.\ 3. Further Fourier
spectrum analysis \cite{cris-unpub} shows that errors actually
become undistinguishable from white noise.

\section{Kinetic roughening picture}
In order to make a theoretical interpretation of our numerical
results, in particular the existence of scaling behavior, we can
resort to the Hopf-Cole transformation, first proposed by Pikovsky
and Politi for the actual Lyapunov vectors \cite{pik1,pik2}, that
allows to link spreading of errors to non-equilibrium surface
growth. We find that the dynamics of finite perturbations can also
be seen as a kinetic roughening processes of the surface defined
by $h(x,t) = \ln \vert \delta u(x,t) \vert$. This is a very useful
transformation that allows us to borrow some well-known results in
the field of nonequilibrium surface growth. Indeed, Politi and
Pikovsky \cite{pik1,pik2} have shown that errors in many extended
systems lead to surface growth process in the universality class
of KPZ \cite{kpz}.

Let us first consider the amplitude factor defined in Eq.\
(\ref{amplitude}) and show that it is related to the average
surface velocity. We can write $\vert \delta u(x,t) \vert = \exp[
h(x,t)]$ and thus $\epsilon(t) = \exp[(1/L)\sum_{x=1}^{x=L}
h(x,t)] = \exp[\overline{h}(t)]$. We then obtain that the
amplitude factor must grow as $\epsilon(t) \sim \epsilon_0
\exp(\lambda_{max} t)$, where $\lambda_{max}$ is the largest
Lyapunov exponent and corresponds in this mapping to the surface
velocity, $\overline{h}(t) = \lambda_{max} t + \overline{h}(0)$
\cite{pik1,pik2}. Direct measures of the average surface velocity
(not shown) are in excellent agreement with our above discussed
estimations of $\lambda$. As an independent check we have also
measured the largest Lyapunov exponent by standard techniques
\cite{eck,geist}.

\begin{figure}
\centerline{\epsfxsize=7.5cm \epsfbox{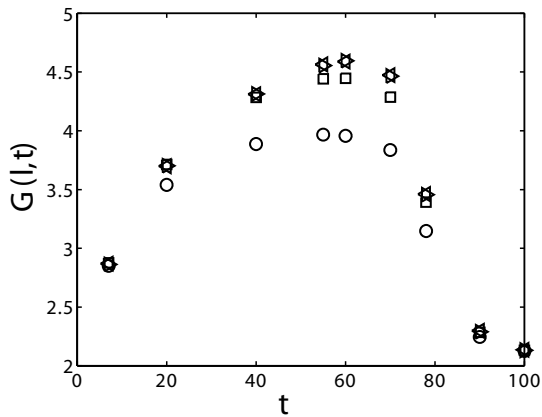}}
\caption{Time behavior of the correlation function for lattice
sites separated at different distances $l=8$ ($\circ$), $l=16$
($\square$), $l=64$ ($\triangleleft$), and $l=185$
($\triangleright$) in an array of $L=1024$ coupled maps.}
\end{figure}

Also the time behavior of the variance of the reduced
perturbations shown in Fig.\ 2 can be linked to surface scaling
properties. From the surface definition $h(x,t) = \ln \vert \delta
u(x,t) \vert$  we have $\sigma^2(t,L) =
\langle\overline{\exp[2y(x,t)]}\rangle$, where $y(x,t) = h(x,t) -
\overline{h}(t)$. We can now proceed and calculate $\sigma^2$
explicitly by considering the cumulant generating function of the
stochastic variable $y$ given by $\Phi(s) = \ln
\langle\overline{\exp(isy)}\rangle$ \cite{gardiner}, where $i$ is
the imaginary unit. This generating function can be expanded in a
power series of the argument $s$ and then evaluated at $s=-2i$ to
obtain an exact expression for the variance $\sigma^2$:
\begin{equation}
\sigma^2(t,L) = \exp[\sum_{r=1}^{\infty} 2^r
{\langle\langle\overline{y^r}\rangle\rangle \over r!}],
\end{equation}
where $\langle\langle\overline{y^r}\rangle\rangle$ are the
cumulants of $y$ \cite{gardiner}. In general, no closed simple
formula exists for the cumulant of order $r$ of a given variable
$y$, but the first cumulants are given by
$\langle\langle\overline{y}\rangle\rangle = \langle\overline{y}
\rangle$, $\langle\langle\overline{y^2}\rangle\rangle =
\langle\overline{y^2}\rangle - \langle\overline{y}\rangle^2$, and
$\langle\langle\overline{y^3}\rangle\rangle =
\langle\overline{y^3}\rangle - 2\langle\overline{y^2}\rangle
\langle\overline{y}\rangle + 2\langle\overline{y}\rangle^3$, etc
\cite{gardiner}. In our case $\langle\overline{y}\rangle = 0$ and
the second cumulant $\langle\langle\overline{y^2}\rangle\rangle =
W^2(t,L)$ corresponds to the surface width $W^2(t,L) = \langle
[\overline{h(x,t) - \overline h]^2}\rangle$. To leading order we
have
\begin{equation}
\label{gaussian}
\sigma^2(t,L) \approx \exp[2W^2(t,L)].
\end{equation}
Higher order cumulants are also finite, but their contribution to
the series gets smaller as $r$ increases. We have checked
numerically that including the 3rd and 4rd cumulants shifts the
exponent in less that a 10\%. Therefore, within this somewhat
crude approximation, Eq.(\ref{gaussian}), we expect to have an
estimation of the {\em effective} scaling exponent of $\sigma^2
\sim \exp(t^{2\beta})$ with less than a 10\% error bar. A
clarification is now in order. For systems in which the largest
Lyapunov exponent $\lambda \to 0$ one would expect to have larger
crossover times $t_{\times}$, so that higher order cumulants may
eventually become relevant, yielding corrections to Eq.\
(\ref{gaussian}). In this sense the scaling picture has to be
considered as approximate. However, the scaling approach is
consistent for hyper-chaotic systems, as those we are interested
here, in which not only the leading Lyapunov exponent is finite,
but a series of finite positive Lyapunov exponents exists and
whose number increases with the system size.

The ansatz of a KPZ behavior for the surface $h$ then implies that
the width should scale as $W(t,L) \sim t^\beta$ for times $t <
t_s(L)$ and saturates to a size dependent value, $W(t,L) \sim
L^\alpha$ for $t> t_s(L)$, where $\beta$ and $\alpha$ are the
growth and roughness exponent, respectively, that are known to
take the values $\beta=1/3$ and $\alpha = 1/2$ for the KPZ
universality class in one dimension \cite{kpz,barabasi}. This is
consistent with the approximate $\exp(t^{2/3})$ scaling observed
in Fig.\ 2, baring in mind the limitation of the approximation
(\ref{gaussian}). It is worth mentioning that all the numerical
results presented here are obtained for system sizes such that
$t_{\times} \ll t_s$, so that the existence of $t_{\times}$ could
be clearly observed.

\section{Conclusions}
We have studied the spatiotemporal dynamics of finite-size
perturbations in extended systems exhibiting chaos. We have
introduced what we call the amplitude factor $\epsilon(t)$ in
Eq.(\ref{amplitude}) and showed that this quantity is directly
related to the maximal Lyapunov exponent and exhibits nice scaling
properties. The amplitude factor allows us to remove in a simple
and self-consistent way the dominant growth component of the
perturbation at every time step, so that spatial correlations can
be determined more easily. At variance with infinitesimal
perturbations, finite-size errors lead to a characteristic time
scale $t_{\times}(\epsilon_0)$ signaling the departure of the
finite perturbation from the linear approach. For times
$t>t_{\times}$ the evolution is nonlinear and spatial correlations
are destroyed. We have found that the perturbation variance grows
as $\sigma(t) \sim \exp(t^{2/3})$, once the contribution of the
largest growth rate is removed. We have explained this behavior by
representing the error growth as a kinetically rough 1D KPZ
surface.

The interpretation of the exponentially growing errors $\delta
u(x,t)$ as a roughening surface $h(x,t) = \ln \vert\delta
u(x,t)\vert$ has many advantages since it allows to translate the
problem of analyzing deterministic chaotic fluctuations into the
simpler framework of kinetic roughening of the associated surface
$h(x,t)$. The scale-invariant character of the surface
fluctuations leads to simple power-law behavior of the relevant
quantities that can be characterized by a few critical exponents.
As a byproduct this leads in a rather simple way to an estimation
of the spatial extent of correlations at time $t$ as $\xi \sim
t^z$, where $z=3/2$ is the dynamic exponent of the KPZ problem.
Further results of practical application, including the relevance
for the predictability problem of the propagation of finite-size
errors and bred vectors, as used in the context of weather
forecasting \cite{kalnay,francisco}, will be published elsewhere
\cite{cris-unpub}.

\begin{acknowledgements}
This work was supported by the CICYT (Spain) through Grants No.
BFM2000-0628-C03-02 and No. BFM2003-07749-C05-03 as well as the EU
Commission through Grant No. OCCULT IST-2000-29683.

\end{acknowledgements}

\end{document}